# Integrated Artificial Neural Network with Trainable Activation Function Enabled by Topological Insulator-based Spin-Orbit Torque Devices

Puyang Huang[1#], Xinqi Liu[2,3#], Yue Xin[1#], Yu Gu[1], A|bert Lee[4], Zhuo Xu[1], Peng Chen[1], Yu Zhang[5], Weijie Deng[2], Guoqiang Yu[5], Di Wu[4], Zhongkai Liu[2,3], Qi Yao[2,3], Yumeng Yang[1], Zhifeng Zhu[1*], and Xufeng Kou[1,3]*

[1]School of Information Science and Technology, ShanghaiTech University, Shanghai 201210, China.

[2]School of Physical Science and Technology, ShanghaiTech University, Shanghai 201210, China.

[3]ShanghaiTech Laboratory for Topological Physics, ShanghaiTech University, Shanghai 201210, China.

[4] Suzhou Inston Technology Co., Ltd., Jiangsu 215121, China.

[5]Beijing National Laboratory for Condensed Matter, Physics Institute of Physics, Chinese Academy of Sciences, Beijing 100190, China.

*Correspondence to: zhuzhf@shanghaitech.edu.cn; kouxf@shanghaitech.edu.cn.

**Non-volatile memristors offer a salient platform for artificial neural network (ANN), but the integration of different function blocks into one hardware system remains challenging. Here we demonstrate the implementation of brain-like synaptic (SOT-S) and neuronal (SOT-N) functions in the $Bi_2Te_3/CrTe_2$ heterostructure-based spin-orbit torque (SOT) device. The SOT-S unit exhibits highly linear (linearity error < 4.19%) and symmetrical long-term potentiation/depression process, resulting in better performance compared to other memristor synapses. Meanwhile, the Sigmoid-shape transition curve inherited in the SOT-N cell replaces the software-based activation function block, hence reducing the system complexity. On this basis, we employ a serial-connected, voltage-mode sensing ANN architecture to enhance the vector-matrix multiplication signal strength with low reading error of 0.61%. Furthermore, the trainable activation function of SOT-N enables the integrated SOT-ANN to execute the Batch Normalization algorithm and activation operation within one clock cycle, which bring about improved on/off-chip training performance close to the ideal baseline.**

Artificial intelligence (AI)-driven digital transformation in modern society requires high-speed and energy-efficient computing[1]. Drawing inspiration from neuroscience, neuromorphic engineering aims to emulate the exceptional computational capabilities of the human brain on solid-state platforms[2,3]. As a key application of this approach, the artificial neural network (ANN) employs weighted synapse units to encode the strengths of the input signals at a connection, and introduces the interconnected neuron units for the input-to-output mapping[4,5]. Therefore, the performance of the synapse unit depends on the generation of numerous intermediate states with high thermal stability, and the corresponding long-term potentiation (LTP) and long-term depression (LTD) process should be linear and symmetric for promoting the learning process[6-9]. Meanwhile, a tunable non-linear activation function with a wide operating range is essential for the neuronal cell component to capture different features of input patterns (**Fig. 1a**)[10-12].

Currently, various ANNs have been realized using conventional digital systems[13-15]. Yet in spite of high classification accuracy, the use of binary bits for large-scale convolution operations is unavoidably time- and energy-consuming[16,17]. Alternatively, memristor-based analog circuits, which naturally support the multiplication and accumulation of electrical signals in linear time-invariant (LTI) systems according to Ohm's and Kirchhoff's laws, can perform the vector-matrix multiplication (VMM) function in parallel. Besides, the results obtained after each convolution operation can be directly stored in the memristor crossbar array, eliminating the need for data transfer[18,19]. Like other analogue ANN paradigms[18,20-25], the non-linear and non-volatile spin dynamics can also empower magneto-resistive devices with memristor-like behaviors, hence enabling energy-efficient neuromorphic computing[26,27]. For example, dedicated multi-layer structural engineering can tailor the mesoscopic spin-texture and anisotropic energy of magnetic domains, which serve as the intermediate state variables of an artificial synapse[28-32]. Meanwhile, the non-destructive electrical manipulation of magnetic domains through spin-transfer torque (STT) and/or spin-orbit torque (SOT) would give rise to a symmetric resistance change with high programming linearity and endurance (*i.e.*, essential for realizing the synapse function with high accuracy)[33]. Furthermore, it is found that the transition curve during the current-induced magnetization switching process exhibits a Sigmoid-featured line-shape; in this regard, the SOT device itself can also fulfill the artificial neuron function without using additional activation function modules in peripheral circuits[28]. Along with programmable logic and non-volatile storage capabilities in the SOT-based magneto-resistive random-access memory (SOT-MRAM)[34], the all-in-one analogue ANN computing system, in principle, could be developed based on the SOT-type devices.

To harness the aforementioned properties, in this Article, we fabricate the SOT-based devices on the 2-inch $Bi_2Te_3/CrTe_2$ heterostructures. We demonstrate reliable SOT-driven multi-state magnetization switching and showcase the basic operating principles of brain-like synapse (SOT-S) and neuron (SOT-N) units in crossbar arrays. The linear and symmetric weight response to the input current pulses with evenly spaced intermediate states of SOT-S and intrinsic Sigmoid-shape transition curve of SOT-N ensure higher classification accuracy and a simpler system layout compared to other memristors. To take full advantage of the SOT-ANN framework, we introduce serial-connected synapses with a voltage-mode sensing architecture to enhance the output VMM signal strength while reducing circuit complexity. Additionally, we utilize the thickness-dependent ferromagnetism of the $CrTe_2$ layer to optimize both the LTP/LTD learning process and the activation function. Furthermore, we implement the Batch Normalization (BN) algorithm in our SOT-ANN using trainable SOT-N units to achieve improved training results.

**Implementing the SOT-Synapse and SOT-Neuron units.** Experimentally, high-quality van der Waals (vdW) $Bi_2Te_3/CrTe_2$ heterostructures were prepared on the 2-inch $Al_2O_3$ (0001) substrate using molecular beam epitaxy (MBE), with detailed sample information provided in Methods and Supplementary Fig. S1. In this bilayer stack illustrated in **Fig. 1b**, the strong SOC-related spin-momentum-locking mechanism of the topological surface states ensures an effective spin polarization from the $Bi_2Te_3$ layer[35-37]. In the meantime, the inherent perpendicular magnetic anisotropy (PMA) of $CrTe_2$ allows for direct pairing with the adjacent $Bi_2Te_3$ channel without invoking additional PMA-assisted layer, which in turn yields a greatly enhanced SOT efficiency of $\xi_{SOT} > 1$ compared to conventional heavy metals[38]. Consequently, the deterministic SOT-driven magnetization switching was realized in our $Bi_2Te_3$(18 nm)/$CrTe_2$(21 ML)-based crossbar devices with a low switching current density of $J_{SW} = 2.84 \times 10^6$ A/cm$^2$ at $T = 120$ K (Supplementary Fig. S2). Moreover, due to the non-destructive SOT switching mechanism, the recorded $R_{xy}$(HIGH)-to-$R_{xy}$(LOW) contour in **Fig. 1c** does not experience any distortion after $5 \times 10^4$ write/read cycles. Furthermore, the device-to-device variation in terms of $J_{SW}$ is found to be less than 4.69% across the array (**Fig. 1d**), thanks to the uniform morphology of the as-grown $Bi_2Te_3/CrTe_2$ wafer[39].

Strikingly, the multi-domain nature of the $CrTe_2$ layer results in the emergence of multiple intermediate state variables during magnetization switching in the fabricated $Bi_2Te_3$(18 nm)/$CrTe_2$(21 ML) SOT-device, and the resulting minor loops of current-modulated $V_{xy}$-$I_{DC}$ in **Fig. 1e** exhibit two distinctive features. Firstly, changing the write current pulse ($I_{\mathrm{write}}^{\mathrm{syn}}$) successively from 15 mA to 27 mA generates 12 repeatable Hall resistance states, representing the LTP process of the SOT-S cell with a 3-bit weight. Similarly, by reversing the write current pulse polarity, a symmetric and linear LTD slope is produced, as shown in **Fig. 1f**. In view of the SOT-S performance, the LTP/LTD curves from 6 trials of identical potentiation/depression training pulses all maintain high linearity (*i.e.*, the average linearity error is less than 4.19%, as shown in Supplementary Fig. S3) and low write stochasticity (*i.e.*, standard deviation of write stochasticity in **Fig. 1g** is $\sigma = 5.41\%$). Moreover, after updating weight information in the SOT-S device, the resulting magnetic domain structure becomes stable against thermal fluctuation. As highlighted in **Figs. 1h-1i**, the readout Hall voltages remain constant as long as the read current range (*i.e.*, 0.1 mA $\leq I_{\mathrm{read}}^{\mathrm{syn}} \leq$ 10 mA) is below $I_{\mathrm{write}}^{\mathrm{syn}}$, and the corresponding read stochasticity is only 0.37% (*i.e.*, the reading error analysis is provided in Supplementary Fig. S4). Besides, **Fig. 1j** shows that the measured $V_{\mathrm{read}}^{\mathrm{syn}}$- $I_{\mathrm{read}}^{\mathrm{syn}}$ slopes at all weights have a high-linearity feature in the $I_{\mathrm{read}}^{\mathrm{syn}} \subseteq [-10\ \mathrm{mA}, 0\ \mathrm{mA}]$ region, hence enabling a wide operating range of the SOT-S unit with low reading discrepancy (Supplementary Fig. S5). Concurrently, the magnetization switching curves of the $Bi_2Te_3/CrTe_2$-based crossbar device can also fulfil the SOT-Neuron function (*i.e.*, in contrast, other memristor-based ANN need additional digital modules or load resistors[40]). As depicted in **Fig. 1e**, all $V_{xy}$-$I_{DC}$ minor-loops in the positive [0 mA, 27 mA] region can be well-fitted by the logistic function as $y = [1 + e^{-k(x-x_c)}]^{-1}$ with $k$ = 0.89 and $x_c$ = 17.59 (**Fig. 1k**); on the other hand, the transition line-shape and relevant ($k$, $x_c$) values in the negative $I_{DC}$ range are adjustable according to the write current level ($I_{\text{write}}^{\text{neuron}}$), as discussed later. In conclusion, the aforementioned characterizations manifest the feasibility to integrate both SOT-S (*i.e.*, for weight encoding) and SOT-N (*i.e.*, for activation function) units on one chip, making them compelling building blocks for a more efficient neural network.

**Constructing the SOT-based integrated neural network.** According to the operational principle of the SOT crossbar device, we adopted the serial-connected synapses architecture with voltage-mode sensing to construct the integrated ANN for the standard Modified National Institute of Standards and Technology (MNIST) test (**Fig. 2a**). In this specific design illustrated in **Fig. 2b**, the serial-connected SOT-S array requires only one analog-to-digital converter (ADC) per row, and it also enables direct superposition of the SOT-S output voltage signals so that the peripheral sensing circuit design can be simplified[28]. In addition, since both the write and read currents share the same conduction path in the SOT-S unit, the number of digital-to-analog converter (DAC) thus is halved to further reduce the circuit complexity[41].

To conduct the MNIST test using our $Bi_2Te_3/CrTe_2$-based crossbar devices, the reset current $I_{\text{reset}}^{\text{syn}}$ is firstly applied along the **-*y*** direction to reset all SOT-S cells to the initial magnetized state (*i.e.*, normalized $R_{xy}$ = 0). Afterwards, the programmed write current pulse $I_{\text{write,ij}}^{\text{syn}}$ (*i.e.*, which exerts appropriate SOT to tune the device to the required intermediate state) flows through each synapse node $S_{\text{ij}}$ to update the weight information of $R_{xy,\text{ij}}^{\text{syn}}$ (**Fig. 2c**). Once the weight assignment is completed, the input current $I_{\text{VMM,ij}}^{\text{syn}}$ is converted into the output Hall voltage signal of $V_{\text{VMM,ij}}^{\text{syn}} = I_{\text{VMM,ij}}^{\text{syn}} \cdot R_{xy,\text{ij}}^{\text{syn}}$ via anomalous Hall effect (AHE), and the VMM operation is accomplished by successively summing the SOT-S cells along the same row as $V_{\text{VMM,i}}^{\text{syn}} = \sum_j V_{\text{VMM,ij}}^{\text{syn}}$. Finally, such output voltage is delivered into the SOT-N cell (upper right panel of **Fig. 2c**) where the activation function of the SOT-driven switching curve is utilized to complete the training process (the operation modes of our ANN are discussed in Supplementary Section 6). As a proof-of-concept, we fabricated a serial-connected synapse array consisting of three SOT-S devices ($S_1$, $S_2$, $S_3$), as shown in the bottom panel of **Fig. 2c**. When the input currents ($I_{\mathrm{VMM},1}^{\mathrm{syn}}, I_{\mathrm{VMM},2}^{\mathrm{syn}}, I_{\mathrm{VMM},3}^{\mathrm{syn}}$) are applied, the measured output voltage $V_{\mathrm{out}}^{\mathrm{syn}}$ is highly consistent with the ideal scenario of $V_{\mathrm{VMM}}^{\mathrm{syn}} \equiv I_{\mathrm{VMM},1}^{\mathrm{syn}} \cdot R_{xy,S1}^{\mathrm{syn}} + I_{\mathrm{VMM},2}^{\mathrm{syn}} \cdot R_{xy,S2}^{\mathrm{syn}} + I_{\mathrm{VMM},3}^{\mathrm{syn}} \cdot R_{xy,S3}^{\mathrm{syn}}$, where the weight information is set as ($R_{xy,S1}^{\mathrm{syn}}, R_{xy,S2}^{\mathrm{syn}}, R_{xy,S3}^{\mathrm{syn}}$) = (0.39 Ω, 0.44 Ω and 0.32 Ω), respectively (**Fig. 2d**). By mapping all ($I_{\mathrm{VMM},1}^{\mathrm{syn}}, I_{\mathrm{VMM},2}^{\mathrm{syn}}, I_{\mathrm{VMM},3}^{\mathrm{syn}}$) combinations in the [0 mA, 5 mA] range, the maximum error of $|V_{\mathrm{out}}^{\mathrm{syn}} - V_{\mathrm{VMM}}^{\mathrm{syn}}|/V_{\mathrm{VMM}}^{\mathrm{syn}}$ is found to be lower than 0.61% (**Fig. 2e**), thereby validating the high precision of the VMM operation (results of other input cases are exemplified in Supplementary Fig. S7).

Inspired by the above [1 × 3] VMM demonstration, we subsequently employed a three-layer SOT-based ANN for handwritten digit recognition, utilizing off-chip training to quantize the trained digital model and deliver results to the analog neural network[42,43]. Benefiting from a better LTP/LTD process in terms of symmetry, linearity, and state number (**Fig. 2f**), our integrated ANN system achieves a higher classification accuracy of 90.34% than other memristors, as summarized in **Fig. 2g** (*i.e.*, it is noted that we directly adopted experimental data of both SOT-S and SOT-N cells, whereas other networks used software-generated Sigmoid activation function for off-chip training simulations)[28,29,44,45]. To gain further insights into the feature extraction process, we traced the feature evolution through $\sum_{i=1}^{n} W_{1,\mathrm{i}} \times W_{2,\mathrm{ij}}$ ($\mathrm{j} = 1 \sim m$), where $n$ ($m$) is the size of the hidden (output) layer, and $W_{1,\mathrm{i}}$ ($W_{2,\mathrm{ij}}$) is the weight array between the input (hidden) layer and the hidden (output) layer. As visualized in the lower panel of **Fig. 2h** and Supplementary Fig. S8, the characteristics of the 0~9 digits quickly become distinguishable after only 50 iterations (*i.e.*, classification accuracy of 66.7%), and the final results trained by our SOT-S/SOT-N system after 1000 iterations are also of better pattern quality (Supplementary Fig. S9).

**Improving SOT-ANN performance through $CrTe_2$ thickness optimization.** In our previous study, we discovered a thickness-dependent ferromagnetism of $CrTe_2$ (**Fig. 3a**), which provides an additional degree of freedom for manipulating the SOT-S and SOT-N units[46,47]. Specifically, the nearly square-shaped $R_{xy}$ hysteresis loops in **Fig. 3b** confirm the presence of PMA in a series of $CrTe_2$ samples with the film thickness ($d$) from 5 ML to 21 ML. On the other hand, as emphasized in **Fig. 3c**, the coercive field $H_C$ monotonically decreases from 12.6 mT ($d$ = 21 ML) to 7.5 mT ($d$ = 5 ML) at 120 K, leading to an effective lowering of the critical switching current $I_{SW}$ by more than 55% in thinner $Bi_2Te_3/CrTe_2$ films (*i.e.*, corresponding to a 51.2% energy reduction of the SOT-S write operation). More importantly, the measured anomalous Hall resistance $R_{xy}$ of the $Bi_2Te_3$(18 nm)/$CrTe_2$(5 ML) sample increases by 10 times compared with the $Bi_2Te_3$(18 nm)/$CrTe_2$(21 ML) data. Consequently, the enlarged $R_{xy}$ magnitude accommodates 18 intermediate state variables (*i.e.*, 4-bit weight), accompanying by a 6-fold improvement in the resistance margin between neighboring states (**Fig. 3d**) This not only ensures a wider operating range (*i.e.*, higher bit resolution) of the SOT-S cell, but also mitigates the design challenges related to the sensing accuracy of the readout circuit. Likewise, by varying the $CrTe_2$ layer thickness, the $R_{xy}$-$I_{DC}$ transition slopes and their associated ($k$, $x_c$) values are modulated accordingly. **Figure 3e** shows that the most non-linear activation function with largest $k$ = 1.076 is obtained in the $Bi_2Te_3$(18 nm)/$CrTe_2$(5 ML)-based SOT-N device, which help to better capture different features of complex patterns[28]. As a result of this structural engineering strategy, the SOT-ANN developed from the $Bi_2Te_3$(18 nm)/$CrTe_2$(5 ML) heterostructures achieves a higher classification accuracy of 93.45%, as highlighted in **Fig. 3f** (detailed characterizations of the $d$ = 5 ML SOT device are provided in Supplementary Fig. S10). Therefore, the above results justify that the thickness-tailored ferromagnetism of $CrTe_2$ adds flexibility to optimize both the off-chip training and the power dissipation of the neural network.

Remarkably, the improved SOT-S/SOT-N performance in our $Bi_2Te_3$(18 nm)/$CrTe_2$(5 ML) system also allows for on-chip training, which conducts the iteration process in the memristor array to reduce power consumption (*i.e.*, off-chip training utilizes software-based iteration), albeit at the expense of classification accuracy due to finite intermediate states[48,49]. To meet the minimum 8-bit precision requirement for on-chip training[50,51], we chose to link 16 SOT-S cells (each with 18 intermediate states) in our serial-connected network, resulting in a total of 288 available states in one synapse unit cell. Along with the linear and evenly spaced LTP/LTD process of SOT-S which warrants uniformly distributed state variables across the entire weight quantization range, our $d$ = 5 ML SOT-ANN attains a relatively high classification accuracy of 83.15%, as shown in **Fig. 3g**. In contrast, suffering from a worse LTP/LTD linearity, uneven state distribution, and the absence of the bit-operation function, the on-chip training results of other memristor-based systems under the equivalent 8-bit quantization condition are not acceptable for practical applications.

**Integrated Batch Normalization and activation functions enabled by tunable $R_{xy}$-$I_{DC}$ minor-loops.** As neural networks for large-scale model training become more sophisticated, an increased number of hidden layers incurs problems such as gradient explosion or gradient disappearance during back-propagation, due to uneven data distribution in forward propagation[52]. To mitigate such issues, the BN algorithm has been proposed to eliminate internal covariate shift, thereby ensuring a stable input range among iterations (**Fig. 4a**)[10,53]. Mathematically, the BN block transforms the VMM result $x$ to $x'$ through $x' = \gamma x + \beta \equiv BN_{\gamma,\beta}(x)$, where $\gamma$ and $\beta$ represent the scale and shift parameters, respectively[42]. This converted signal $x'$ is then passed to the activation function block as $f(x') = [1 + e^{-k(x'-x_c)}]^{-1} = [1 + e^{-\gamma k(x-\frac{x_c-\beta}{\gamma})}]^{-1}$, which is equivalent to generate a new set of activation function parameters $k' = \gamma k$ and $x_c' = \frac{x_c - \beta}{\gamma}$. While most analog ANN systems conduct BN algorithm by software, the SOT-N cell enables both BN and Activation operations to be realized in a single cell thanks to its tunable ($k$, $x_c$) characteristics. For instance, in the case of the $Bi_2Te_3$(18 nm)/$CrTe_2$(5 ML)-based device, by changing the write current $I_{write}^{neuron}$ from 19 mA to 32 mA, the tuning range of $k$, which is extracted from the normalized $V_{VMM}^{neuron}$-$I_{VMM}^{neuron}$ transition curves of **Fig. 4b**, is found to be [0.37, 1.28]. Consequently, the conventional five-stage analog ANN system can be simplified into a four-stage diagram consisting of VMM, BN&Activation, Back-Propagation, and Weight-Update, with only the Back-Propagation block being performed by software, as illustrated in **Fig. 4c**.

In accordance with this generic SOT-ANN framework, we further designed a relevant multiplexer circuit to synchronize the BN and Activation operations within a single clock cycle. As outlined in **Fig. 4d**, during the BN operation (*i.e.*, the ENACT signal is set as 0), a large negative reset current pulse from DAC initializes the SOT-N, followed by the positive write signal $I_{write}^{neuron}$ to set the SOT-N with the desirable ($k$, $x_c$) values; in the Activation operation (*i.e.*, ENACT = 1), the output voltage $V_{VMM,i}^{syn}$ from the $i^{th}$ row of the SOT-S array is applied onto the SOT-N cell to perform the Sigmoid calculation (right panel of **Fig. 4d**). Accordingly, by incorporating the BN algorithm in our $d$ = 5 ML SOT-ANN, the MNIST classification accuracy of the off-chip training is increased from 93.45% to 95.38%, and the training loss is reduced to 0.273, both of which are close to the ideal benchmark (left panel of **Fig. 4e**). Similarly, compared to the network with fixed activation function, the tunable ($k$, $x_c$)-enabled BN algorithm also helps improve the on-chip training result by 6.18% as well as lower the loss from 1.5 to 0.8 with the same iteration number of 400 (right panel of **Fig. 4e**). Consequently, the co-optimization of hardware and algorithm in our integrated SOT-ANN paves the way for the realization of advanced and versatile AI-related computing.

## Discussion

In conclusion, the investigation of $Bi_2Te_3/CrTe_2$ heterostructure-based SOT devices in this work reveals several advantages, including high linearity, long write/read endurance, and integrated synapse-neuron functionality over other memristor technologies, as summarized in **Table 1**[54-56]. Moreover, by leveraging the trainable activation function by the write current pulse-configurable $V_{\mathrm{VMM}}^{\mathrm{neuron}}$-$I_{\mathrm{VMM}}^{\mathrm{neuron}}$ transition slope, we demonstrate the BN algorithm using the spintronic hardware with improved classification accuracy in both on-chip and off-chip training, which would be beneficial for low-power edge-computing and Internet of Things (IoT) applications[57]. In addition, since SOT-type devices exhibit salient size scaling (< 100 nm) and fast switching (< 5 ns) properties[58], they may offer new opportunities of our proposed $Bi_2Te_3/CrTe_2$-based system for constructing energy-efficient large-scale neuromorphic computing systems[32,33].

## Methods:

***Sample growth and structural characterizations***: Wafer-scale $CrTe_2$ and $Bi_2Te_3/CrTe_2$ thin films were grown on 2-inch $Al_2O_3(0001)$ substrates by MBE under a vacuum of $1 \times 10^{-10}$ mbar. Before sample growth, the $Al_2O_3$ substrate was pre-annealed at 600 °C to remove adsorbed contamination. The substrate temperature was kept at 200 °C during both $CrTe_2$ and $Bi_2Te_3$ growth. High-purity Cr and Bi atoms were evaporated from standard Knudsen cells, and Te was evaporated using a thermal cracker cell. The real-time growth conditions and as-grown surface atom configuration were monitored by reflection high-energy electron diffraction (RHEED) patterns, and the flux ratio was calibrated by in-situ beam flux monitor. After sample growth, slices of $Bi_2Te_3/CrTe_2$ with different crystal orientations were obtained through focused ion beam (FIB) milling using the TESCAN LYRA3 FIB-SEM (TESCAN, Czech Republic). The crystal structure of the samples was analyzed using a probe aberration-corrected scanning transmission electron microscopy (Cs-STEM, Themis Z G2 300, FEI, USA) equipped with energy-dispersive X-ray analysis (EDX, Bruker Super-X, Bruker, USA) to map the element distribution.

***Device fabrication and transport measurements***: The MBE-grown $CrTe_2/Bi_2Te_3$ heterostructures were patterned into μm-sized cross-bar structures using standard photo-lithography, and the Ti(15 nm)/Au(150 nm) electrodes were defined by e-beam evaporation. After device fabrication, magneto-transport measurements were performed in a $He^4$ refrigerator (Oxford Teslatron PT system) using multiple lock-in amplifiers and Keithley source meters. During the SOT-driven magnetization switching experiment, a 500 μs writing current pulse was applied by Keithley 6221, followed by a 500 μs reading current pulse to measure $R_{xy}$ by Keithley 2182. For the [1 × 3] VMM demonstration, three serial-connected SOT-S devices were controlled by three independent Keithley 6221, and the output voltage was recorded 200 ms after the application of input currents to ensure stable output states. Besides, a 5 s delay time was imposed between each operation to minimize unintentional Joule heating.

***Network framework and parameter-setup for simulation***: The image classification of handwritten digits from the MNIST database was simulated using a neural network by Matlab. The proposed three-layer SOT-ANN in this study consisted of 784 (28 × 28) input neurons, 128 hidden neurons, and 10 output neurons. Both off-chip and on-chip training simulations (up to 1000 iterations) were carried out based on 50,000 and 10,000 handwritten digit images as the training and testing dataset, respectively. The activation operation in our SOT-ANN was based on the experimental data of the SOT-N cells, while an ideal Sigmoid function with $k = 1$ and $x_c = 0$ was used for other memristors. The cross-entropy loss function was employed for the back-propagation algorithm in all network systems.

## Acknowledgments

This work is sponsored by the National Key R&D Program of China (grant no. 2021YFA0715503), National Natural Science Foundation of China (grant no. 92164104), and the Major Project of Shanghai Municipal Science and Technology (grant no. 2018SHZDZX02), the Shanghai Engineering Research Center of Energy Efficient and Custom AI IC, and the ShanghaiTech Quantum Device and Soft Matter Nano-fabrication Labs (SMN180827). X.K. acknowledges support from the Shanghai Rising-Star Program (grant no. 21QA1406000) and the Open Fund of State Key Laboratory of Infrared Physics.

## Author contributions

X.K. conceived and supervised the study. X.L. and P.C. grew the samples by MBE. P.H., X.L., and W.D. performed structural characterizations and transport measurements. P.H., X.L., Y.Z., G.Y., Z.L., Q.Y., and Y.Y. analyzed the experimental data. P.H., Y.X. and Z.X. conducted the neural network simulations. P.H., Y.G., A.L., and D.W designed the circuit structure and performed power consumption evaluation. P.H., Z.Z. and X.K. wrote the manuscript. All authors discussed the results and commented on the manuscript.

## Competing financial interests

The authors declare no competing financial interests.

## Figures

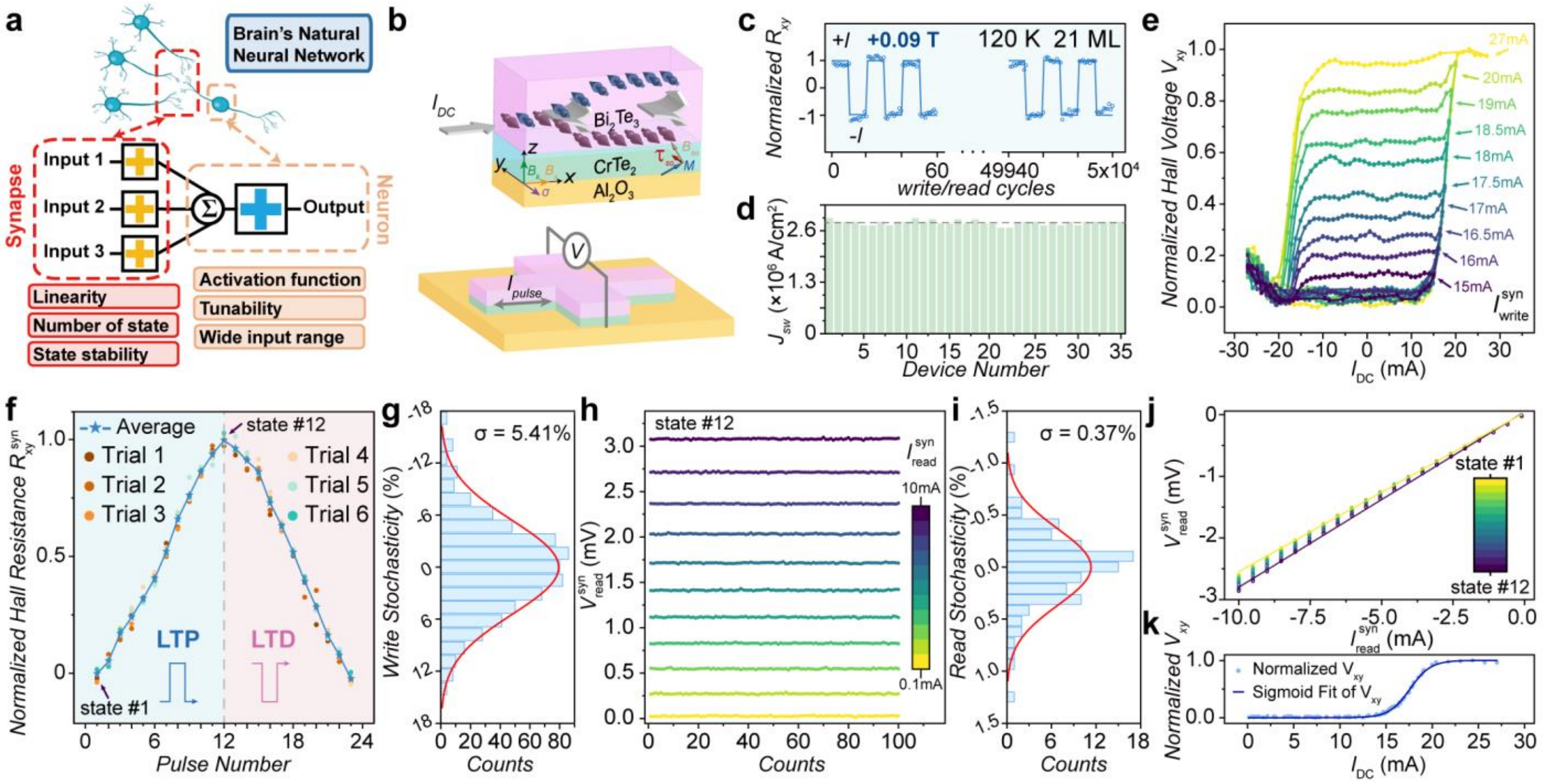

**Fig. 1. Characterizations of $Bi_2Te_3$(18 nm)/$CrTe_2$(21 ML) heterostructure-based SOT-S and SOT-N devices. a,** Schematic of neural network and requirements of the synapse and neuron units for reliable learning and recognition. **b,** Illustration of the SOT-driven magnetization switching mechanism in the $Bi_2Te_3$/$CrTe_2$ heterostructures. **c,** Endurance test of the $Bi_2Te_3$(18 nm)/$CrTe_2$(21 ML) crossbar up to $5 \times 10^4$ write/read cycles at $T$ = 120 K. An in-plane magnetic field of $B$ = 0.09 T was applied to enable the deterministic switching. **d,** Device-to-device variation of 36 SOT devices across the 2-inch wafer. **e,** Memristor-like SOT switching minor-loops of the $Bi_2Te_3$(18 nm)/$CrTe_2$(21 ML) system. 12 intermediate states are generated by $I_{\mathrm{write}}^{\mathrm{syn}}$ ranging from 15 mA to 27mA. **f,** Symmetric and repeatable LTP/LTD curves of the $d$ = 21 ML SOT-S device under 6 trials of training pulses. **g,** Write stochasticity distribution of the SOT-S device where the data were collected from Fig. 1f. **h-i,** The readout Hall voltage $V_{\mathrm{read}}^{\mathrm{syn}}$ (**h**) and corresponding read stochasticity distribution of $I_{\mathrm{read}}^{\mathrm{syn}}$ = 5 mA (**i**) of the SOT-S device. Once the weight information is set by $I_{\mathrm{write}}^{\mathrm{syn}}$, the measured $V_{\mathrm{read}}^{\mathrm{syn}}$ exhibits little fluctuation after 100 counts of the read

current pulses (0.1 mA < $I_{\mathrm{read}}^{\mathrm{syn}}$ < 10 mA) for all 12 state variables. **j,** Linear $V_{\mathrm{read}}^{\mathrm{syn}} - I_{\mathrm{read}}^{\mathrm{syn}}$ slopes of the SOT-S device under different weights. **k,** Fitting result of the Sigmoid-type $V_{xy}$-$I_{\mathrm{DC}}$ transition curves in the positive [0 mA, 27 mA] region by the logistic function.

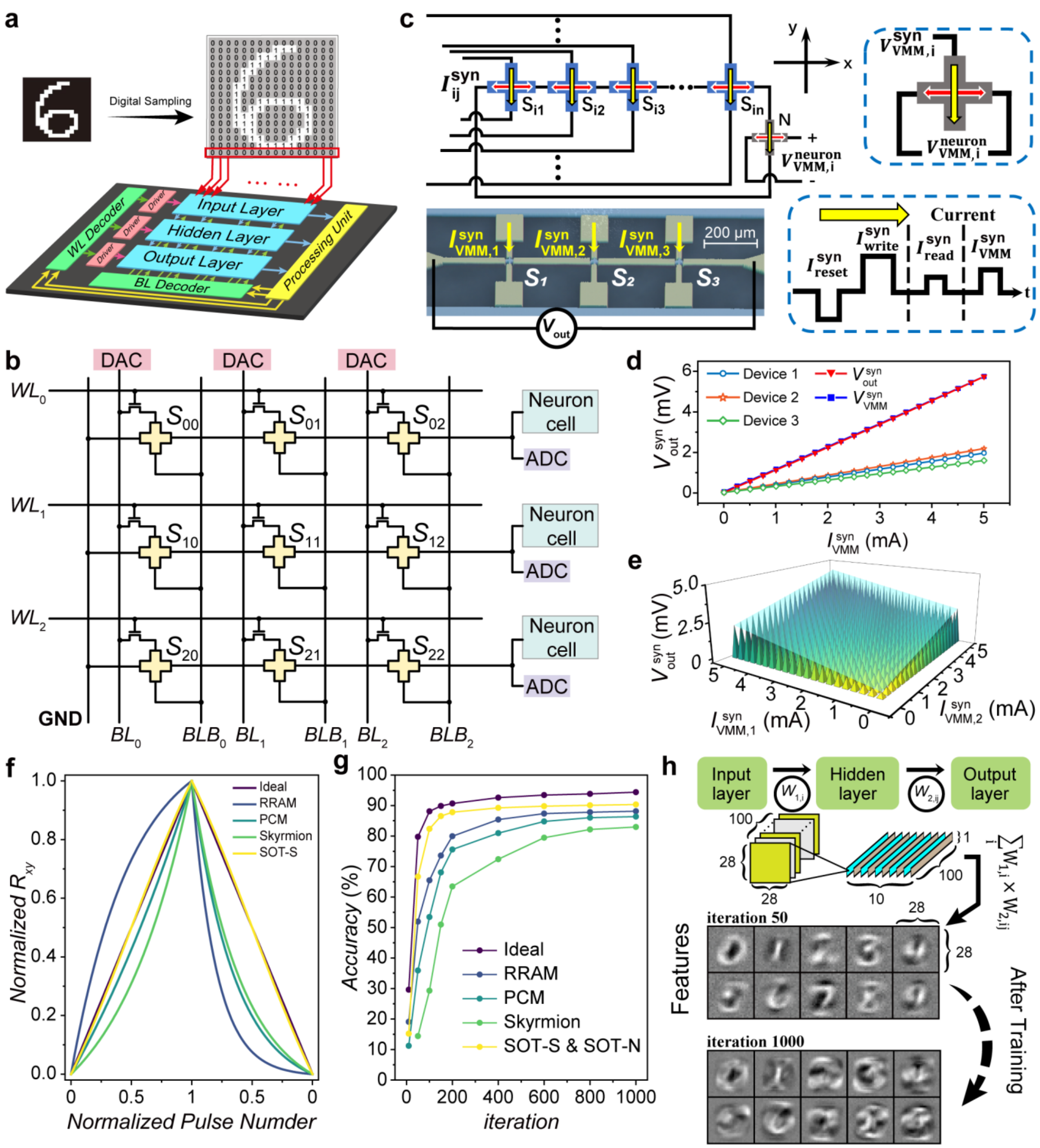

**Fig. 2. Circuit architecture and image classification accuracy of the integrated SOT-ANN. a,** Schematic of the neural network used for the MNIST test. **b,** Serial-connected, voltage-mode sensing SOT-ANN architecture. **c,** Operating principle of the integrated SOT-ANN network. After the initialization ($I_{\mathrm{reset}}^{\mathrm{syn}}$) and weight assignment ($I_{\mathrm{write,ij}}^{\mathrm{syn}}$), the input current ($I_{\mathrm{VMM,ij}}^{\mathrm{syn}}$) is applied through each SOT-S device (–*y* direction), and the output Hall voltage signal ($V_{\mathrm{VMM,ij}}^{\mathrm{syn}}$) is summed along the same row (*x*-axis) which is subsequently served as the input voltage of the SOT-N unit. The lower panel of Fig. 2c displays the microscopic image of three serial-connected *d* = 5 ML SOT-S units ($S_1$, $S_2$, $S_{3.}$) used for the [1 × 3] VMM demonstration, where $I_{\mathrm{VMM,1}}^{\mathrm{syn}}$, $I_{\mathrm{VMM,2}}^{\mathrm{syn}}$ and $I_{\mathrm{VMM,3}}^{\mathrm{syn}}$ represent the input currents. **d**, Measured $V_{\mathrm{out}}^{\mathrm{syn}}$ (red triangles) and ideal $V_{\mathrm{VMM}}^{\mathrm{syn}}$ (blue squares) results of the three serial-connected SOT-S array with the weight information of 0.39 Ω (cyan circles), 0.44 Ω (orange stars) and 0.32 Ω (green diamonds), respectively. **e,** $V_{\mathrm{out}}^{\mathrm{syn}}$ mapping under one specific input current condition with fixed $I_{\mathrm{VMM,3}}^{\mathrm{syn}}$ = 1 mA but varied ($I_{\mathrm{VMM,1}}^{\mathrm{syn}}$, $I_{\mathrm{VMM,2}}^{\mathrm{syn}}$) combinations. **f-g,** Comparison of the normalized LTP/LTD curves (**f**) and the off-chip trained classification accuracy (**g**) among different memristor technologies. **h,** The recognition features of the 0~9 digits after the off-chip training after 50 and 1000 iterations.

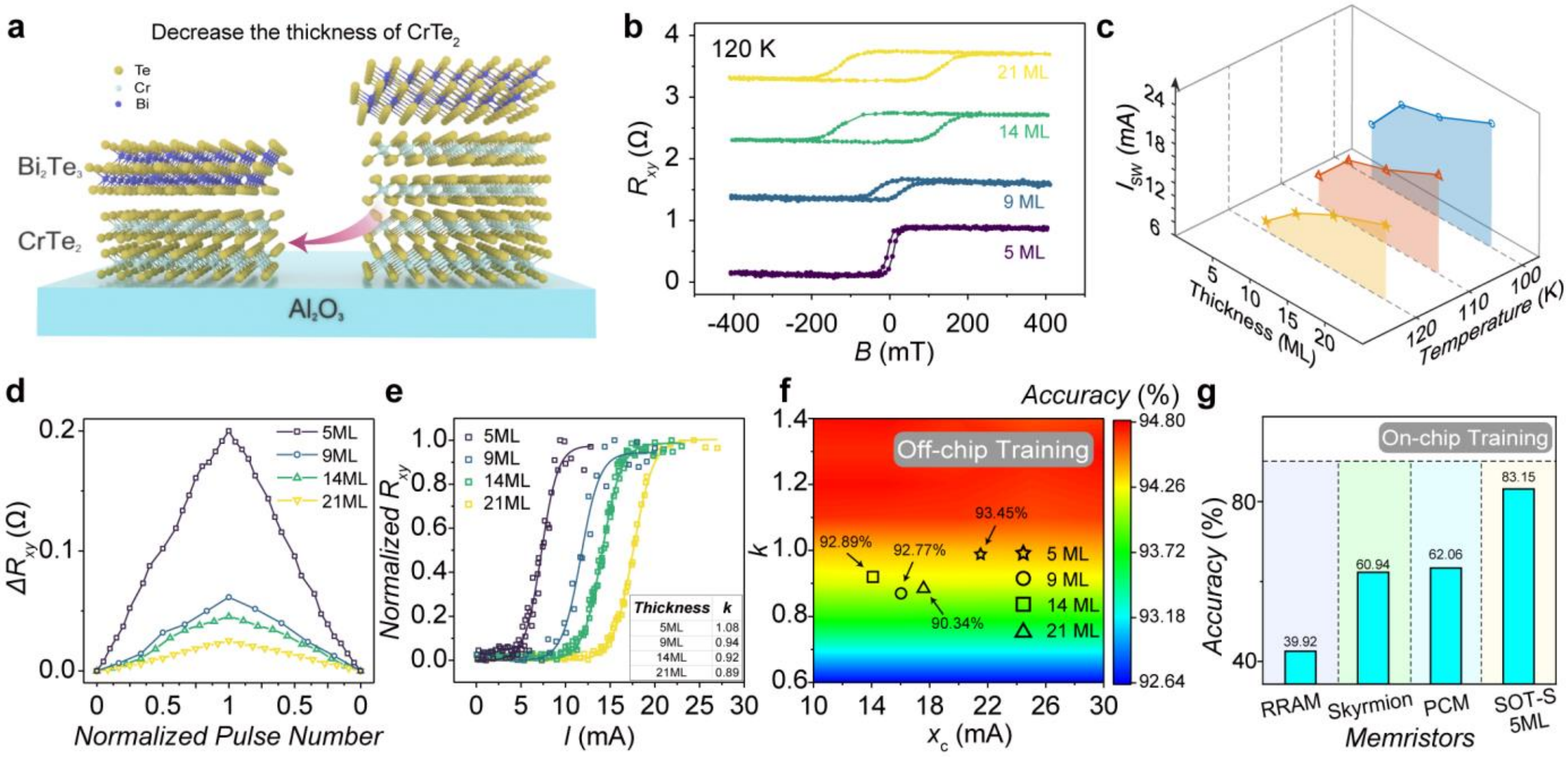

**Fig. 3. Dependence of the SOT-ANN performance on the $CrTe_2$ layer thickness. a,** Schematic of $Bi_2Te_3$/$CrTe_2$ heterostructures with adjustable $CrTe_2$ layer thickness. **b,** Anomalous Hall resistances of four $CrTe_2$ samples with $d$ = 5 ML, 9 ML, 14 ML, and 21 ML, respectively. Data are shifted vertically for convenient comparison. **c,** The evolution of $I_{SW}$ as functions of temperature and $CrTe_2$ layer thickness. **d-e,** Tunable LTP/LTD curves (**d**) and $R_{xy}$-$I_{DC}$ transition slopes (**e**) of the $Bi_2Te_3$(18 nm)/$CrTe_2$($d$)-based SOT devices. **f,** Off-chip training results with different $CrTe_2$ layer thicknesses. The background diagram was conducted under ideal synapse units with infinite weight states. **g,** Comparison of the on-chip trained classification accuracy of different memristors under the same 8-bit quantization condition after 1000 iterations.

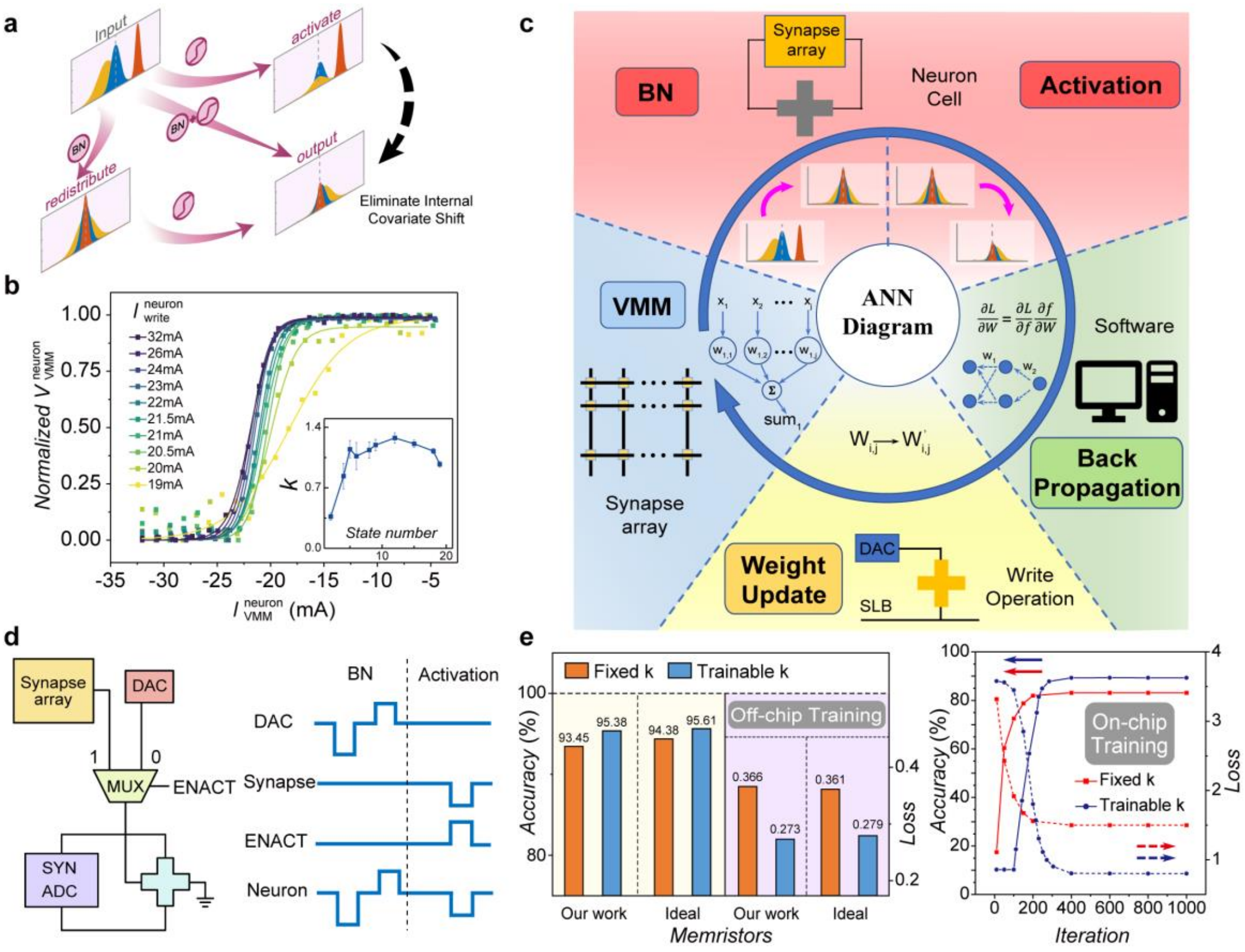

**Fig. 4. Four-stage integrated SOT-ANN system with available Batch Normalization and trainable activation functions. a,** Correction of the internal covariate shift by applying the BN algorithm. **b,** Normalized $V_{\mathrm{VMM}}^{\mathrm{neuron}}$ - $I_{\mathrm{VMM}}^{\mathrm{neuron}}$ transition curves and the extracted $k$ values (**inset**) of the $Bi_2Te_3$(18 nm)/$CrTe_2$(5 ML)-based SOT device. The write current range (*i.e.*, which generates 18 intermediate states) is 19 mA < $I_{\mathrm{write}}^{\mathrm{neuron}}$ < 32 mA. **c,** Illustration of the four-stage SOT-ANN system with the corresponding working units (VMM, BN&Activation, Back-Propagation, and Weight-Update). **d,** The circuit diagram (**left panel**) and operating signal waveforms (**right panel**) to execute both the BN and Activation operations in the same SOT-N device within one clock cycle. It is noted that the output voltage of the SOT-S array $V_{\mathrm{VMM,i}}^{\mathrm{syn}}$ is converted into a negative $I_{\mathrm{VMM}}^{\mathrm{neuron}}$ by setting $I_{\mathrm{VMM}}^{\mathrm{neuron}} = -V_{\mathrm{VMM,i}}^{\mathrm{syn}}/R_{\mathrm{ch}}^{\mathrm{neuron}}$, where $R_{\mathrm{ch}}^{\mathrm{neuron}}$ is the SOT-N channel resistance. **e,** Comparison of the classification accuracy and loss in the $d$ = 5 nm SOT-ANN system with fixed and trainable $k$ after the off-chip (**left panel**) and on-chip (**right panel**) trainings.

| Technology | CMOS Memories | | Memristive Device | | | | |
|---|---|---|---|---|---|---|---|
| | NOR Flash | NAND Flash | RRAM | PCM | FeFET | STT-MRAM | Our Work |
| Linearity | Low | Low | Low | Low | Low | Medium | High |
| Integration density | High | Very High | High | High | High | High | Medium |
| Switching Energy | High | High | Medium | High | Medium | Low | Low |
| Retention | Long | Long | Medium | Long | Long | Medium | Long |
| Endurance | $10^5$ | $10^4$ | $10^5$-$10^8$ | $10^6$-$10^9$ | $>10^5$ | $10^{15}$ | $>10^{15}$ |
| Inherited Activation Function | Yes | Yes | No | No | No | No | Yes |
| BN Algorithm & Activation Function integration | Yes | Yes | No | No | No | No | Yes |

**Table 1. Summary of analog ANN performance implemented by different memristor platforms.**